\documentclass[aps,pra,twocolumn,final]{revtex4-1}

\usepackage{graphicx}
\graphicspath{{/Users/trensink/Dropbox/publication/LatexFigures/images/}{/Users/trensink/Desktop/LatexFigures/images/}{/Users/trensink/Desktop/FINAL_IMAGES/images/}}
\usepackage{amssymb} 
\usepackage{amsmath} 
\usepackage{mathrsfs}
\usepackage[usenames]{color} 
\usepackage{textcase}
\usepackage{bm}
\usepackage{subfigure}
\usepackage{setspace}
\usepackage{natbib}
\newcommand{\ignore}[1]{}
\begin{document}

\title{Model for the Atomic Dielectric Response in Time Dependent Laser Fields}

\author{T.C. Rensink}\affiliation{University of Maryland, College Park}
\author{T.M. Antonsen Jr.}\affiliation{University of Maryland, College Park}
\author{J.P. Palastro}\affiliation{University of Maryland, College Park}
\author{D. Gordon}
\affiliation{Naval Research Laboratory}

\date{11/21/2013}

\begin{abstract}
A nonlocal quantum model is presented for calculating the atomic dielectric response to a strong laser electric field.  By replacing the Coulomb potential with a nonlocal potential in the Schrodinger equation, a 3+1D calculation of the time-dependent electric dipole moment can be replaced with a 0+1D integral equation, offering significant computational savings. The model is benchmarked against an established ionization model and \textit{ab initio} simulation of the time-dependent Schrodinger equation.  The reduced computational overhead makes the model a promising candidate to incorporate full quantum mechanical time dynamics in laser pulse propagation simulations.

\end{abstract}

\maketitle	

\section{INTRODUCTION}

High intensity ultrashort laser propagation gives rise to a wide range of phenomena and has been an integral part of several fields of research over the past few decades such as laser wakefield acceleration \citep{Malka2013}, generation of terahertz radiation \citep{Johnson2013, Kim2007}, high harmonic generation \citep{Krause1992}, and atmospheric filamentation \citep{Couairon2007, Chen2010, Wahlstrand2011, Palastro2012}.  For laser intensities near the ionization threshold of tenuous propagation media, field and medium dynamics become strongly nonlinear in the electric field, requiring numerical simulation for proper treatment.  In principle, this would be accomplished by calculating the self-consistent evolution of all fields and charges in the system.  Consider for a moment a collection of non-interacting atoms of Hydrogen.  For a single such atom with center of mass at position $\mathbf R$, we can express the atomic dipole moment in terms of the relative coordinate, $\mathbf r' \equiv \mathbf r - \mathbf R$:
 \begin{equation}\label{dipole}
\mathbf d(\mathbf R, t) \equiv -e \langle \mathbf r' \rangle. 
\end{equation}
where $\langle \mathbf r' \rangle \equiv \int \mathrm{d^3\mathbf r'} \ \psi^*(\mathbf r', t) \mathbf r' \psi(\mathbf r',t)$ is the expectation value of the electron position relative to $\mathbf R$ in terms of the electron wavefunction, $\psi(\mathbf r',t)$.  On summing the dielectric contribution over all of the atoms and weighting by the local gas density $n_g(\mathbf R)$, we arrive at the macroscopic polarization density 
\begin{equation}\label{pnl}
\mathbf P(\mathbf R,t) =  n_g(\mathbf R) \mathbf d(\mathbf R, t)
\end{equation}
that appears as a source term in a propagation equation for the laser electric field, $\mathbf E(\mathbf R,t)$.  In practice, simulating laser pulse propagation over macroscopic distances is complicated by the need to include quantum mechanical dynamics at atomic scales; while calculations of this kind have been performed for single atoms with the time dependent Schrodinger equation (TDSE), such simulations generally require supercomputing resources.  Scaling this up to a full 3D laser propagation simulation is a task well beyond current capabilities.

Instead, the ``standard'' treatment of material dielectric response in propagation simulations consists of breaking the total polarization density in Eq.\eqref{pnl} into separate terms for various limiting cases.  In the limit that the laser field is small compared with the atomic field, a ``bound'' atomic response is given by a perturbative expansion in the laser field strength.  Associated polarization density terms are proportional to powers of the laser electric field strength, where, for example, $\mathbf P^{(1)} \propto \mathbf E$ and $\mathbf P^{(3)} \propto |\mathbf E|^2 \mathbf E$.  A separate plasma response term is used to account for the ``free'' electrons that have been ionized.  Because ionization is often dominated by quantum tunneling, a separate physical model must be introduced to calculate the ionization rate.  A final term must be included to account for the field energy lost during ionization.

The shortcomings of this treatment are reflected in the fact that several terms are necessary to describe the total nonlinear polarization density given in Eq.\eqref{pnl}.  While \textit{ab initio} quantum simulations aren't possible for the reason stated above, we explore a nonlocal interaction (NLI) model \citep{TetchouNganso2011} that offers a promising alternative: by replacing the Coulomb potential with a nonlocal potential term in the 3D TDSE, a computationally inexpensive method can be used to calculate $\mathbf d(\mathbf R, t)$ for a single atom, putting a non-fragmented treatment of $\mathbf P(\mathbf R, t)$ for laser propagation simulations within reach.  Our objectives for this paper are then twofold: present the NLI model, and motivate its validity.  Sections \ref{sec:formulation} and \ref{sec:solution} introduce the formulation of the NLI model and explain the method used for fast computation.  Section \ref{sec:system} gives expressions for the ionization rate, bound electron probability, and expectation value of the electron position that are used to benchmark the NLI model in section \ref{sec:benchmark}.  Finally, section \ref{sec:extensions} briefly outlines an extension to include an arbitrary number of bound states.  Other detailed analyses are deferred to the appendix to permit continuity of the main text.

\section{FORMULATION}\label{sec:formulation}
We consider a single atom of Hydrogen at the origin, such that $\mathbf R = 0$ and $\mathbf r' = \mathbf r$ (the convention used hereafter).  The 3D TDSE for a single electron under the influence of a Coulomb potential, $-Z/\mathbf r$, and subject to a time varying electric field $ \mathbf E(t) $ in the dipole approximation is: 
\begin{equation}\label{TDSE}
i \hbar \frac{\partial}{\partial t}\psi(\mathbf r,t) = \left[ -\frac{\hbar^2}{2m_e}\nabla^2 - \frac{Ze^2}{|\mathbf r|} + e\mathbf E(t)  \cdot \mathbf r  \right]\psi(\mathbf r,t).
\end{equation}
The nonlocal interaction potential (NLI) model is formulated by replacing the Coulomb potential in Eq.\eqref{TDSE} with a nonlocal potential term:
\begin{equation*} \label{eq:coulomb}
-\frac{Ze^2}{|\mathbf r|}\psi(\mathbf r,t) \to -Vu(\mathbf r)S(t),
\end{equation*}
where
\begin{subequations}
\begin{align}
u(\mathbf r) &\equiv \left(\pi \sigma^2 \right)^{-3/4} \exp\left(-\mathbf r^2/2\sigma^2 \right), \label{udef}\\
S(t) &\equiv \! \int \! \! \mathrm{d^3\mathbf r} \ u(\mathbf r)\psi(\mathbf r,t).\label{sdef}
\end{align}
\end{subequations}
Here $u(\mathbf r)$ gives the spatial extent of the binding potential, while $S(t)$ is the ``nonlocal'' portion of the potential.  Using the shorthand notation $\langle f(\mathbf r) | g(\mathbf r) \rangle  \equiv \int \mathrm{d^3 \mathbf r} \, f(\mathbf r) g(\mathbf r)$, it can be seen from Eq.\eqref{udef} that $u(\mathbf r)$ is a normalized function such that $|\langle u| u\rangle|^2 = 1$ and $S(t) = \langle u(\mathbf r) | \psi(\mathbf r, t)\rangle $ is the projection of the time-dependent wavefunction onto $u(\mathbf r)$.  We have introduced two free parameters with the above definitions: $V$, the normalized binding energy, used to change the overall strength of the binding potential, and $\sigma$, the spatial extent of the binding potential.  These may be chosen to match the ionization properties of atomic Hydrogen, as will be seen later.

It is not immediately apparent that this is an appropriate replacement for the Coulomb potential, and a few comments are in order to motivate this substitution.  The nonlocal nature of the modified potential prevents it from being drawn on energy-space axes as can be done for local potentials.  However, in the limit $\sigma \rightarrow 0$, $u( \mathbf r)$ becomes a delta function and the NLI potential reduces to a more familiar local potential term, $-\lambda \delta^3(\mathbf r)\psi(\mathbf r,t)$; while the delta potential has been considered in 1D treatments of Eq.\eqref{TDSE} \citep{Teleki2010}, the 3D extension produces solutions to $\psi(\mathbf r)$ that are singular at the origin.  The NLI potential can be considered an extension of the 3D delta function that permits normalizable solutions to the wavefunction.  An important feature of the NLI model is that it maintains the hermitian, unitary, and linear properties of Eq.\eqref{TDSE}.  

For simplicity, we introduce the following normalized quantities that will be used in the remainder of the paper: $\mathbf r/\sigma \to  \mathbf r$, $\hbar t/m_e \sigma^2  \rightarrow t$, $m_e \sigma^2 V/\hbar^2 \rightarrow V$, and $\mathbf \sigma^3 m_e e\mathbf E(t)/\hbar^2  \rightarrow  \mathbf E(t).$  These are identical to atomic units except that all factors of length are scaled to $\sigma \equiv \beta a_0$ ($\beta$ constant) instead of the Bohr radius.  Although normalized quantities depend on the as-yet unspecified value for $\sigma$, this normalization simplifies the algebraic expressions considerably.  Below are the previously defined quantities in the normalized coordinates, as well as the modified TDSE.  Together, these form the complete system of equations we wish to solve for a general electric field, $\mathbf E(t)$:
\begin{subequations}\label{system}
\begin{align}
u(\mathbf r) &= \pi ^{-3/4} \exp \left(-\mathbf r^2/2\right) \label{u},\\
S(t) &=  \! \int \! \mathrm{d^3\mathbf r} \: u(\mathbf r)\psi(\mathbf r,t) \label{Sdef} ,\\
i \frac{\partial}{ \partial t}\psi(\mathbf r,t) &= \left[ -\frac{1}{2}\nabla^2 + \mathbf r \cdot \mathbf E(t) \right]\psi(\mathbf r, t) - Vu(\mathbf r)S(t)  \label{NLITDSE}.
\end{align}
\end{subequations}

\section{SOLUTION}\label{sec:solution}
In principle, the system given by Eqs.\eqref{system} could be simulated directly by time evolving the modified Schrodinger equation with a finite-volume \citep{Kono1997} or spectral method \citep{Bauer2006}, and quantities of interest could be obtained through the usual prescription of operators and expectation values.  This would be a computational task essentially equal to solving the original TDSE, with no advantage gained by using the modified binding potential.  However, the NLI model offers a considerably different approach to obtain the same information.  Specifically, we reduce the system given by Eqs.\eqref{system} to an integral equation in time for $S(t)$ without explicitly calculating $\psi(\mathbf r,t)$.  Quantities of interest, such as the dipole moment $\mathbf d(t)$ and the bound probability of the electron can, in turn, be derived directly in terms of $S(t)$, thereby eliminating the need to solve for the wavefunction altogether.

The computational savings of the NLI model are a direct result of the fact that no spatial representation for $\psi(\mathbf r,t)$ is required to be calculated to obtain information about the system.  By contrast, a typical finite volume treatment of the TDSE calculates $\psi(\mathbf r ,t)$ on a spatial grid and evolves it at each point in space over time.  Accurate calculation of $\langle \mathbf r \rangle$ requires the spatial domain to be large enough to capture free-wavefunction excursions on the order of the quiver radius $r_{q} = e|E_L|/m_e \omega_L^2$ (where $\omega_L$ and $E_L$ are the frequency and amplitude of the applied field repectively) while maintaining sufficient spatial resolution to resolve the wavefunction of large momentum states.  The time domain must resolve the period of the quantum bound state (typically sub-femtosecond), while extending over the duration of the laser pulse simulation, often on the order of hundreds of femtoseconds.  While still subject to the same time domain constraints, the NLI approach lifts the restrictions in the spatial domain entirely, as will be seen.

A Green's function (or propagator) approach will be used to obtain the integral equation for $S(t)$.  We first define $G(\mathbf r,t;t')$ as the solution to the equation 
\begin{equation}
\left[ i \frac{\partial}{ \partial t} +\frac{1}{2}\nabla^2 - \mathbf r \cdot \mathbf E(t) \right]G(\mathbf{r},t;t')  = i u(\mathbf r)\delta(t-t'),\label{LG}
\end{equation}
where we have taken Eq.\eqref{NLITDSE} and replaced $-VS(t)$ by an impulse in the time domain, $i \delta(t-t')$, and the boundary condition is taken to be $G(\mathbf r, t<t';t') = 0$.  Because the electric potential term $-\mathbf r \cdot \mathbf E(t)$ is linear in space, Eq.\eqref{LG} admits a closed form solution, 
\begin{equation} \label{Gdef}
\begin{split}
G(&\mathbf r,t;t') =  \frac{1}{\pi^{3/4} \left[ 1 + i(t-t')\right]^{3/2}}\times \\ 
&\exp \left[ i \mathscr S_0 +i \mathbf v_0 \!  \cdot \!|\mathbf r - \mathbf r_0|  - \frac{|\mathbf r - \mathbf r_0|^2}{2+2i(t-t')}  \right],
\end{split}
\end{equation}
The function $G(\mathbf r,t;t')$ depends on the trajectories of a classical electron, designated with subscript ``o'', subject to field $\mathbf E(t)$.  Here, $ \mathbf r_0(t;t')$, $\mathbf v_0(t;t')$, and $\mathscr S_0(t;t')$, represent the position, velocity, and action of a classical electron, related through the coupled ordinary differential equations with associated initial conditions:
\begin{subequations}\label{traj}
\begin{align}
\mathrm{\frac{d \mathbf r_0 }{d t}} &=  \mathbf v_0(t),\label{traj1}\\
\mathrm{\frac{d \mathbf v_0 }{d t}} &= -\mathbf E(t),\label{traj2}\\
\mathscr S_0(t,t') &\equiv \int _{t'}^t \mathrm{dt''} \left[ \;\tfrac{1}{2} \mathbf v_0^2(t'') + \mathbf r_0(t'') \cdot \mathbf E(t'') \right], \label{traj3} \\
\intertext{ where} 
\mathbf v_0(t&=t';t') = \mathbf r_0(t=t';t') = 0. \label{traj4}
\end{align}
\end{subequations}
Conceptually, these trajectories describe the path of an electron ``born'' at the origin with zero initial velocity at $t'$ and subsequently moving under the force of the electric field until $t$.  With the function $G(\mathbf r, t;t')$ defined by Eq.\eqref{LG} and Eq.\eqref{Gdef} we can express the wavefunction as a convolution,
\begin{equation} \label{psigf}
\psi(\mathbf r,t) = iV \!\! \int \limits_{-\infty} ^t \mathrm{dt'} G(\mathbf r, t;t')S(t').
\end{equation}

To make use of this expression for $\psi(\mathbf r,t)$, $S(t')$ must be known on the interval $-\infty < t' \leq t$.  For problems of interest, we will assume that the wavefunction is in the bound state and $\mathbf E(t) = 0$ for $t \leq 0$ - this constraint is sufficient to obtain an analytic expression for $S(t')$ on this domain.  For $t>0$ (after the field is present) values of $S(t')$ must be calculated with a general expression for $S(t)$, obtained in the following way: on inserting Eq.\eqref{psigf} into Eq.\eqref{Sdef} and integrating over all space, an integral equation in time for $S(t)$ for general field $\mathbf E(t)$ is given by
\begin{equation}\label{Sgen}
S(t) =  i2^{3/2}V \!\! \int \limits_{-\infty}^t \mathrm{dt'} S(t') \frac{\exp\left[ i \mathscr S_0(t,t') + \Lambda(t,t') \right]}{\left[ 2+ i(t-t') \right]^{3/2}},
\end{equation}
where
\begin{equation*}
\Lambda(t,t') \equiv \frac{1 + i(t-t')}{2+i(t-t')} \frac{|\mathbf r_0 - i \mathbf v_0|^2}{2}-\frac{1}{2}\mathbf r_0^2. 
\end{equation*}
Equation \eqref{Sgen} will be used to calculate the time dependence of all quantities of interest, including the dipole moment and $\psi(\mathbf r, t)$ via Eq.\eqref{psigf}, and as such is the primary computational task in the NLI model.  The time savings over typical TDSE treatments is manifest by the absence of any spatial dependence in Eq.\eqref{Sgen}.  One might protest that we've traded the problem of a large spatial simulation domain for an infinite time integral, but solving for $S(t)$ via Eq.\eqref{Sgen} is more tractable than it might seem: as with Eq.\eqref{psigf}, the explicit form of $S(t')$ on $-\infty < t' \leq 0$ is obtained with the condition that the electron is bound on this interval, while subsequent values of $S(t)$ can be calculated numerically via Eq.\eqref{Sgen}.  A more detailed discussion of the numerical treatment of Eq.\eqref{Sgen} is given in the appendix. 

\section{SYSTEM PROPERTIES}\label{sec:system}
\subsection{Field Free System}
\begin{figure*}
\centering
{\includegraphics{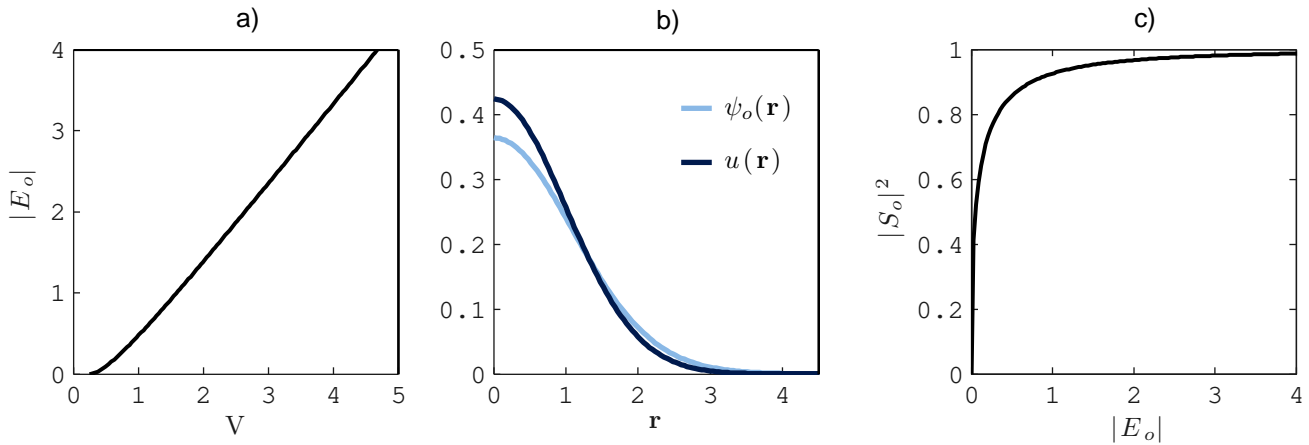}}
\caption{a) Normalized bound state energy $E_0$ as a function of normalized potential strength $V$.  Sufficiently small $V$ does not admit a bound state. Figure 1. \!\!b) Profiles for the bound state wavefunction $\psi_0(\mathbf r)$ and the NLI function $u(\mathbf r)$ for parameter values $V = 3.77$, $\sigma = 2.494 a_0$ used for modeling atomic Hydrogen.  Figure 1. \!\!c) The quantity $|S_0|^2 = |\langle u(\mathbf r)| \psi_0(\mathbf r) \rangle |^2$ as a function of normalized bound state energy: $\psi_0(\mathbf r)$ approaches $u(\mathbf r)$ in the high energy limit.}
\label{fig:dispersion}
\end{figure*}

To better understand the nonlocal potential, we first examine the system in the absence of an applied field.  For $\mathbf E(t) = 0$, the NLI potential admits a single bound state $\psi_0(\mathbf r)$ with energy $E_0$ that can be determined as follows.  With no applied field, classical variables $\mathbf r_0(t,t')$, $\mathbf v_0(t,t')$ and $\mathscr{S}_0(t,t')$ in Eqs.\eqref{traj} are identically zero, and Eq.\eqref{Sgen} simplifies to a convolution whose kernel depends only on the time difference $(t-t')$,
\begin{equation}\label{So}
S(t) = iV2^{3/2}\int \limits_{-\infty}^t \mathrm{dt'} \frac{ S(t')}{\left[ 2 + i(t-t') \right]^{3/2}}.
\end{equation}
Solutions of Eq.\eqref{So} are of the form $S(t) = S_0e^{-iE_0t}$, where $S_0$ is a complex constant.  Inserting this expression into Eq.\eqref{So} with $\mathbf E(t) = 0$ results in a transcendental equation for the energy $E_0$ given by
\begin{equation}\label{dispersion}
V=\frac{1}{4}\left[1 - \sqrt{2\pi|E_0|}e^{2|E_0|}\mathrm{erfc}(\sqrt{2|E_0|}) \right]^{-1}.
\end{equation}
The expression in Eq.\eqref{dispersion} is plotted in Fig.\ref{fig:dispersion} \!\!a).  Sufficiently large values of $V$ correspond to a single bound state wavefunction of the form $\psi(\mathbf r,t) = \psi_0(\mathbf r ) e^{-iE_0t}$ and eigenvalue $E_0$.  An expression for the bound state wavefunction can be found by inserting Eq.\eqref{Gdef} into Eq.\eqref{psigf} with $\mathbf r_0, \mathbf v_0, \mathscr S_0 = 0$ and $S(t) = S_0e^{-iE_0t}$ to give
\begin{equation}\label{psi0}
\psi_0(\mathbf r) = i\frac{S_0V}{ \pi^{3/4}} \int \limits^\infty_0 \mathrm{dt'} \: \frac{e^{iE_0t'}}{\left[1+it'\right]^{3/2}}\exp\left[{\frac{-|\mathbf r|^2}{2+2it'}}\right].
\end{equation}
The profile of $\psi_0(\mathbf r)$ is plotted in Fig.\ref{fig:dispersion} \!\! b) alongside $u(\mathbf r)$ for comparison.

\subsection{Dipole Moment}
With the ultimate goal of finding the polarization density in mind, we seek a computationally efficient expression for the atomic dipole $\mathbf d(t) = - \langle \mathbf r \rangle$ expressed in terms of $S(t)$ without explicit reference to $\psi(\mathbf r, t)$.  We start with the definitions of the expectation values for normalized position and momentum,  
\begin{align}
\langle \mathbf r \rangle &\equiv \int \mathrm{d^3 \mathbf r} \ \psi^*(\mathbf r, t) \, \mathbf  r \, \psi(\mathbf r, t), \label{position} \\
\intertext{and}
\langle \mathbf p \rangle &\equiv  -i \!\!\int \mathrm{d^3 \mathbf r} \ \psi^*(\mathbf r, t)  \nabla \psi(\mathbf r, t),
\end{align}
noting that both $\langle \mathbf r \rangle$ and $\langle \mathbf p \rangle$ are real.  A set of ordinary differential equations relating $\langle \mathbf r \rangle$ and $\langle \mathbf p \rangle$ is obtained by following the steps used to derive the Ehrenfest relations.  The result is similar to Eqs.\eqref{traj} with additional terms resulting from the nonlocal binding potential:
\begin{subequations}\label{qtraj}
\begin{align}
\frac{\partial \langle \mathbf r \rangle}{\partial t} &= \langle \mathbf {p} \rangle + 2\:\text{Im}\left[S(t) \nabla \!_0 S^*(t)\right], \label{dipole} \\
\intertext{and}
\frac{\partial \langle \mathbf {p} \rangle}{\partial t} &= -\mathbf E(t) + 2\:\text{Re}\left[S(t) \nabla\!_0 S^*(t)\right],\\
\intertext{where}
\begin{split}
\nabla\!_0 S(t) &\equiv -i2^{3/2}V \int \limits_{-\infty}^t \mathrm{dt'} S(t') \ \frac{\left[\mathbf r_0 + i\mathbf v_0 - \mathbf v_0(t-t')\right]}{\left[2+i(t-t')\right]^{5/2}}\times \\
\exp&\left[ i \mathscr S_0(t,t') + \Lambda(t,t') \right].
\end{split}
\end{align}
\end{subequations}
Here, $\Lambda(t,t')$, $\mathscr S_0(t,t')$ are the same as in Eq.\eqref{Sgen}.  After $S(t)$ has been found via Eq.\eqref{Sgen}, Eqs.\eqref{qtraj} can be integrated in time to compute $\mathbf d(t)$.  We note that calculating the dipole moment in this way is computationally more efficient than representing the wavefunction on a grid or with basis modes and computing $\mathbf d(t)$ directly with Eq.\eqref{position}.

\subsection{Bound Probability and Ionization Rate}
Many theoretical \citep{Keldysh1964} and experimental \citep{Walsh1994} studies have focused on laser induced ionization of gases.  For a single electron, a time dependent measure of the bound probability is given by the projection of $\psi(\mathbf r ,t)$ onto the electron's ground state wavefunction, $\psi_0(\mathbf r)$:
\begin{equation}
\rho_b(t) = |\langle \psi_0(\mathbf r) | \psi(\mathbf r, t) \rangle |^2\label{rhob}.
\end{equation}
The bound probability is related to the ionization rate $w$ through the relation:
\begin{align}
\rho_b(t) &= \rho_b(t_0) \exp \left[- \int_{t_0}^t \mathrm{d}t' \ w(t')  \right],\label{bprob}\\
 \intertext{ or} 
w(t) &\equiv -\frac{\partial}{\partial t} \ln\left[ {\rho_b(t)/\rho_b(t_0)}\right].
\end{align}
While the exact expression for the bound probability can be obtained for the NLI model by inserting Eqs.\eqref{psigf} and \eqref{psi0} into Eq.\eqref{rhob}, the result is a cumbersome double time integral.  Instead, a useful proxy for the bound probability is given by projecting onto the normalized function $u(\mathbf r)$:
\begin{equation}
\tilde \rho_b(t) \equiv \frac{|\langle u(\mathbf r) | \psi(\mathbf r ,t) \rangle |^2}{| \langle u( \mathbf r) | \psi(\mathbf r, t= 0) \rangle |^2 }  = \frac{|S(t)|^2}{|S(0)|^2}\label{rhobproxy}
\end{equation}
with associated ionization rate,
\begin{equation}
\tilde w(t) \equiv -\frac{\partial}{\partial t} \ln\left[ {\tilde \rho_b(t)}\right] = -\frac{\partial}{\partial t} \ln\left[\frac{|S(t)|^2}{|S(0)|^2}\right]\label{wproxy}.
\end{equation}
The values of $\rho_b$ and $\tilde \rho_b$ (and hence $w$, $\tilde w$) are exactly equal when the electron is entirely bound, i.e., before the laser field is introduced.  As the atom undergoes ionization, some fraction of the bound electron wavefunction will transition to the continuum of free states, and projecting $\psi(\mathbf r,t)$ onto $u(\mathbf r)$ and $\psi(\mathbf r)$ won't in general give the same result.  After the pulse has passed, free components of the wavefunction continue to spread out in space, leaving only the bound wavefunction near the origin; free state contributions to $\tilde \rho_b$ are accordingly reduced, and $\tilde \rho_b \to \rho_b$ in the long time limit.
\ignore{\begin{figure}
\includegraphics[width=8cm]{PROJ}
\caption{The profiles for the bound state wavefunction $\psi_0(\mathbf r)$ and the NLI function $u(\mathbf r)$ for parameter values $V = 3.77$ (corresponding to $E_0 = -13.6$ eV) and  $\sigma = 2.494 a_0$.}
\label{fig:u_psi}
\end{figure}}

The extent that Eq.\eqref{rhobproxy} represents a reasonable proxy for the bound probability when free wavefunction components are near the origin depends on the similarity of the spatial profiles of $u(\mathbf r)$ and $\psi_0(\mathbf r)$.  For parameters used to model atomic Hydrogen, the spatial profiles of $u(\mathbf r)$ and $\psi_0(\mathbf r)$ are compared in Fig.\ref{fig:dispersion} b).  The similarity in spatial profiles motivates the use of $\tilde \rho_b$, and a direct comparison of $\rho_b$ and $\tilde \rho_b$ in the next section confirms that $\tilde \rho_b$ quite accurately measures the bound probability.  Moreover, the accuracy of $\tilde \rho_b$ is improved for larger values of the bound state energy; if $E_0$ is increased (achieved by increasing $V$), the bound wavefunction $\psi_0(\mathbf r)$ more closely conforms to $u(\mathbf r)$, and $|S_0|^2 = |\langle u(\mathbf r)| \psi_0(\mathbf r) \rangle |^2$ approaches unity, as seen in Fig.\ref{fig:dispersion} c).  In the limit $E_0 \to \infty$, $\tilde \rho_b = \rho_b$ exactly.  For the profile shown in Fig.\ref{fig:dispersion} b), $|\langle u(\mathbf r)| \psi_0(\mathbf r) \rangle|^2 \approx.98$.  

\ignore{
\begin{figure}
\includegraphics[width=7cm]{upsi0}
\caption{}
\label{fig:upsi0}
\end{figure}
}

\section{Modeling Hydrogen}\label{sec:benchmark}

Having presented some of the basic properties of the NLI model, we now attempt to simulate atomic Hydrogen.  In particular, we would like to replicate the dipole response and ionization properties of atomic Hydrogen for typical laboratory ultrashort laser pulse parameters.  

We proceed by comparing simulations of the NLI model against two established models; an \textit{ab initio} TDSE simulation, and a modified version of the well known Keldysh ionization model \citep{Keldysh1964}.  Comparison to each of these provides a different type of validation.  The \textit{ab initio} simulation \citep{Gordon2012} numerically simulates the electron wavefunction time evolution via Eq.\eqref{TDSE}, and provides the highest fidelity treatment of the system we consider in this paper.  While such a comparison is invaluable, the computational demands of full TDSE simulations allow a limited number of runs for comparison.  To investigate the accuracy of the NLI over a range of different parameters, we turn to the Popruzhenko, Mur, Popov, and Bauer (PMPB) \citep{Popruzhenko2008} ionization rate model.  In contrast to the TDSE simulation, the PMPB model does not simulate the time dynamics of the electron wavefunction.  Rather, it only predicts the atomic ionization rate for a monochromatic electric field.  While the PMPB model offers significantly less information than a full TDSE simulation, it can be used to validate the NLI model over a large range of laser frequencies and intensities in relatively short computation time.

To compare the NLI model with those mentioned above, values must be determined for $V$ and $\sigma$.  The value for $V$ was determined by Eq.\eqref{dispersion} such that $E_0 = -13.6$ eV; in as far as we wish to simulate Hydrogen, this is the only choice.  This was modified only slightly in the case of the \textit{ab initio} simulation to match the numerical ground state eigenenergy.  The value of $\sigma$ was determined by matching the total drop in bound probability of the NLI model with that of the \textit{ab initio} simulation (Fig.\ref{fig:g_boundprob}), and used for all comparisons in this paper.

\subsection{\textit{ab initio} TDSE Simulation Comparison}
A brief summary of the the \textit{ab intio} TDSE simulation is as follows: the Coulomb potential in Eq.\eqref{TDSE} is replaced with a soft-core potential, $|\mathbf r|^{-1} \rightarrow (|\mathbf r|^2 + \delta \mathbf r^2)^{-1/2}$, where $\delta \mathbf r = .05$ a.u. is a small constant to accommodate the divergent Coulomb potential on a finite spatial grid.  The Schrodinger equation is put into conservative form and $\psi(\mathbf r,t)$ is propagated via the finite volume method.  The spatial domain consists of $4096 \times 32,768$ (r $\times$ z) cells of size $0.04 \times 0.04$ a.u., and the time domain consists of $40,000 \times 0.04$ a.u. time steps (approximately 40 fs).  Use of the soft-core potential and finite spatial resolution results in a similar eigenspectrum to Hydrogen for the first several bound states.  The numerical ground state energy is slightly displaced from the true energy, corresponding to $E_0 = -13.385$ eV.

In the simulation, a single Hydrogen atom initially in the ground state is subjected to a 14.2 fs (fwhm) linearly polarized laser pulse of 800 nm light with a maximum intensity $\text{I}_{\text{max}} = 2.12\times10^{14} \text{ W/cm}^2$.  The exact form of the field is $E(t) \equiv -\partial A_L/\partial t$ with $A_L(t) = A_0 \sin^2 \! \left( \pi t/\tau \right)  \cos(\omega_L t)$, where $A_0 = 1.37$, $\omega_L = .057$, and $\tau = 800$ in atomic units.

For the \textit{ab initio} run, we define a measure of the bound probability as the electron probability density integrated out to a radius $|\mathbf r| = r_b$:
\begin{equation}\label{rhobproxy2}
\rho_b'(t) \equiv \frac{\int \limits_{|\mathbf r| < r_b } \!\! \mathrm{d^3 \mathbf r} \ |\psi(\mathbf r, t)|^2 }  { \int \limits_{|\mathbf r| < r_b } \!\! \mathrm{d^3 \mathbf r} \ |\psi(\mathbf r, 0)|^2 }. 
\end{equation}

\begin{figure}
\centering
{\includegraphics[width=8.5cm]{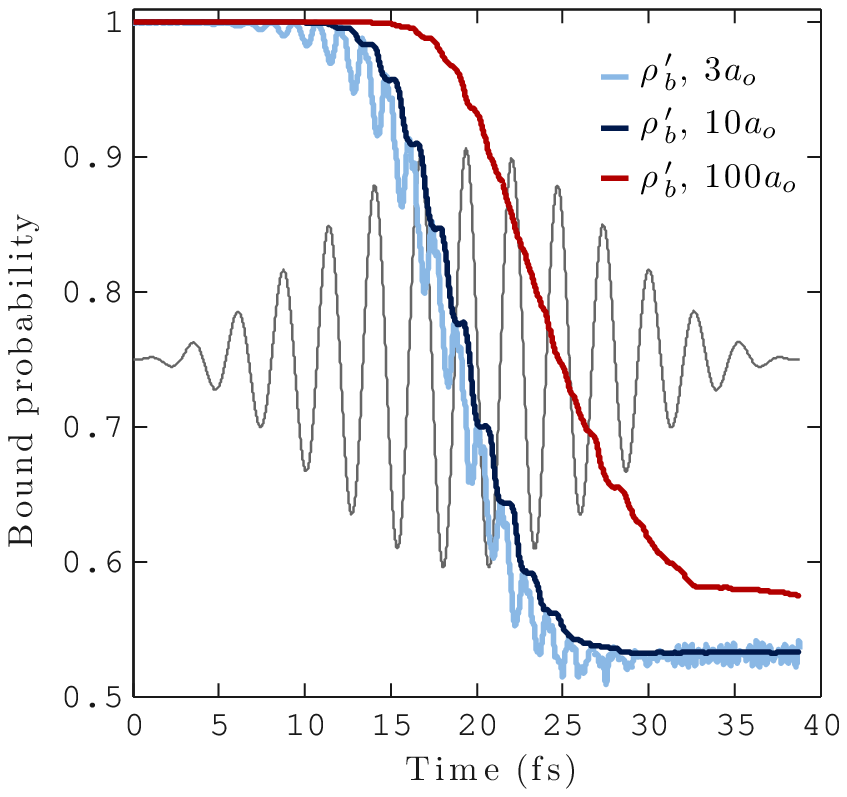}}
\caption{Measures of the bound probability given by Eq.\eqref{rhobproxy2} for three different radii in the \textit{ab initio} simulation normalized to unity.}
\label{fig:gboundprobs}
\end{figure}

The quantity $\rho_b'(t)$ was calculated for three separate radii, $r_b = 3, 10$, and $100$ a.u, with initial integrated probabilities $.934$, $.9999994$, and $\sim 1$ respectively.  Like the approximate bound probability for the NLI model, $\tilde \rho_b(t)$, these values are normalized to unity.  Result are shown in Fig.\ref{fig:gboundprobs}, with the time dependence of $\mathbf E(t)$ included for reference.  

The local minima in the $3 a_0$ curve occurring at peaks in the electric field are a result of the distorted potential well.  As the applied field translates the minimum of the potential well, the ``bound" portion of the wavefunction shifts against the fixed integration region, resulting in the observed minima.  While this feature is largely absent in the $r_b = 10 a_0$ curve, both curves have approximately the same value by the end of the simulation, indicating that approximately 47\% of the wavefunction has transitioned to free states and has dispersed beyond $r_b = 10a_0$.  For the $r_b = 100a_0$ curve however, the probability is still decreasing at 40 fs, indicating that free components of the wavefunction are still propagating out of the integration region.

\begin{figure}
\centering
\includegraphics[width=8.5cm]{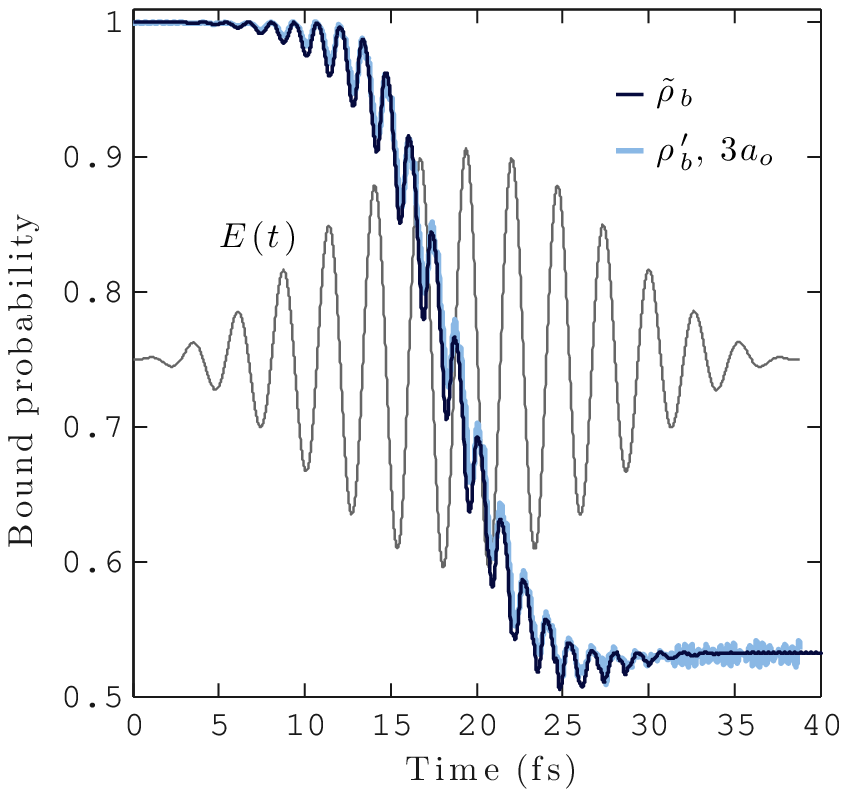}
\caption{A comparison of the $r_b  = 3 a_0$ integrated probability with the approximate bound probability $\tilde \rho_b(t)$ given by the NLI model.}
\label{fig:g_boundprob}
\end{figure}

The analogous run was performed with the NLI model and is compared in Fig.\ref{fig:g_boundprob}.  Here, the overall drop in $\tilde \rho_b$ was matched to the final value of the $r_b = 3a_0$ curve by adjusting the free parameter $\sigma=2.494a_0$, and fixed at this value for all subsequent comparisons (including the PMPB model).  The result shows remarkable agreement in the time dependent structure of $\tilde \rho_b(t)$ and $\rho_b'(t)$. 

\begin{figure}
\subfigure{\includegraphics[width=8cm]{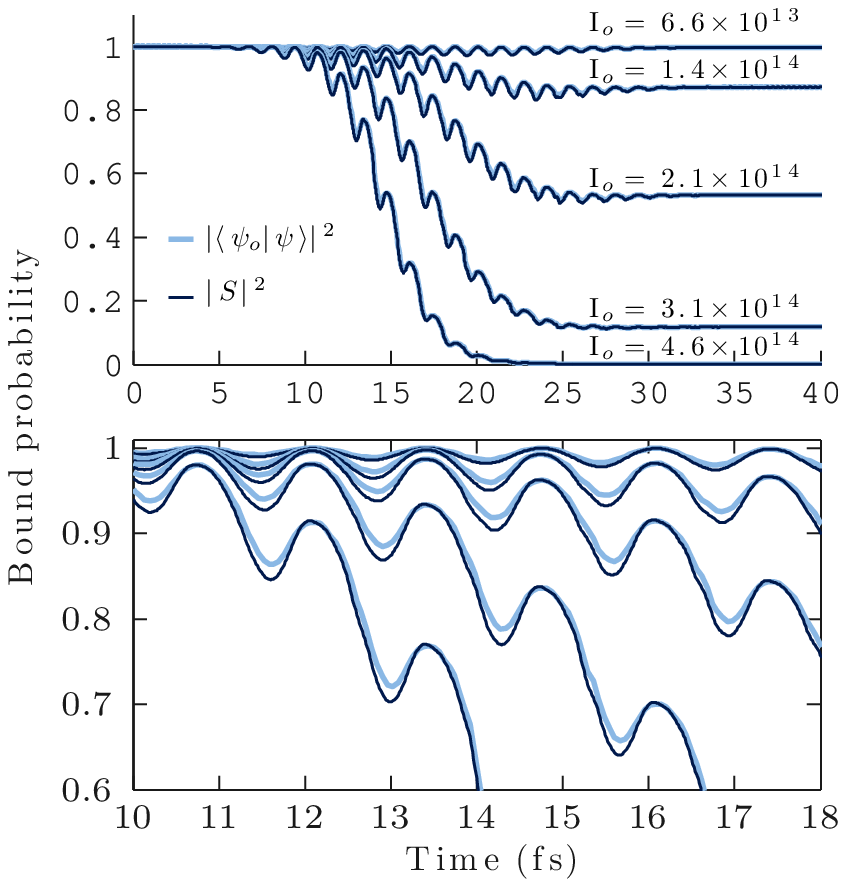}}
\caption{A comparison of $|S|^2$ with the "true" bound probability as calculated by Eq.\eqref{rhob}.  Here, the NLI wavefunction $\psi(\mathbf r,t)$ was calculated via Eq.\eqref{psigf} and integrated to find the total bound probability.  This is compared directly to $|S|^2$ for the same field used in the \textit{ab initio} simulation, with five different peak laser intensities as marked.  \ignore{The above results imply that for the parameters used in this paper, $\tilde \rho_b(t)$ given by Eq.\eqref{rhobproxy} is a reliable measure of the bound probability, independent of the total ionization.}}
\label{fig:integrated}
\end{figure}

Although this agreement is quite suggestive by itself, it is worth examining how accurately $\tilde \rho_b$ represents $\rho_b$ as defined in Eq.\eqref{rhob} for the NLI model.  Figure \ref{fig:integrated} compares $\tilde \rho_b(t)$ and $\rho_b(t)$ for several runs of various laser intensity.  To find $\rho_b$, $\psi_0(\mathbf r)$ and $\psi(\mathbf r,t)$ were calculated via Eqs.\eqref{psigf} and \eqref{psi0} on a spatial grid in $r_\perp$ and $z$.  The resulting probability density was numerically integrated and plotted with $\tilde \rho_b(t)$ in Fig.\ref{fig:integrated}.  For the data provided, the accuracy of $\rho_b(t)$ is limited by integrating on a spatial grid and truncation of the integral in Eq.\eqref{psigf}.  This plot demonstrates the advantage of the NLI model; despite a $\sim 1000\times$ increase in computation time required to solve for $\psi(\mathbf r,t)$ and integrate the result, $\rho_b(t)$ gives a similar results to that of $\tilde \rho_b(t)$.

\begin{figure}
\centering
\includegraphics[width=9cm]{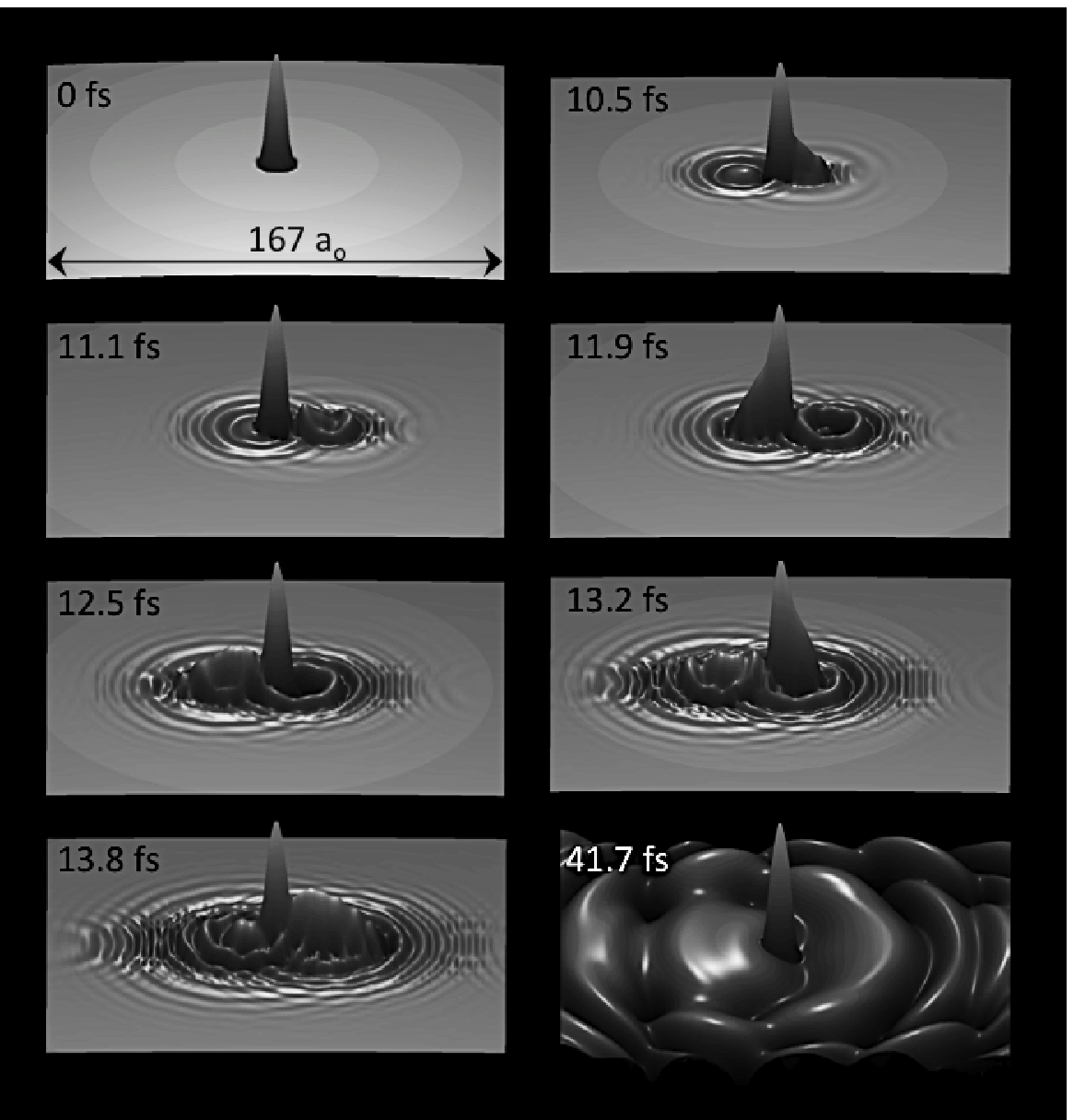}
\caption{The NLI wavefunction density in the $r_\perp \times z$  plane ($167 a_0 \times 167 a_0$), plotted on a natural log scale.  The frames depict the bound state profile (t=0), approximately one laser cycle of evolution (10.5 - 13.8 fs), and a frame shortly after the laser pulse has passed. }
\label{fig:wavefunction}
\end{figure}

Nevertheless, rendering $\psi( \mathbf r,t)$ can be an aid to understanding the time dynamics of the system.  Figure \ref{fig:wavefunction} shows a time series of the NLI electron wavefunction responding to $\mathbf E(t)$ (Fig.\ref{fig:g_boundprob}).  Here, the probability density $|\psi(\mathbf r,t)|^2$ is calculated in the $r_\perp \times z $ plane and plotted on a natural log scale.  The first pane (0 fs) shows the bound state probability density profile, followed by six frames spanning approximately one laser cycle from 10.5 to 13.8 fs, and the last frame shows the wavefunction shortly after the pulse has passed.  The free components still in view at 41.7 fs do not contribute significantly to the bound probability (see Fig.\ref{fig:g_boundprob}) and continue to disperse from the region as time progresses.  For these simulation parameters, the ionized wavefunction continues to interact with the binding potential over the course of the laser period, and interference patterns in the free wavefunction are observed.  Such effects are accounted for in the NLI quantities $\tilde \rho_b$ and $\langle \mathbf r(t) \rangle$, and are not included in simplified rate models.  For this simulation, $\tilde \rho_b$ and $\langle \mathbf r(t) \rangle$ can be calculated in less than a minute.

\begin{figure}
\centering{}
\includegraphics[width=8cm]{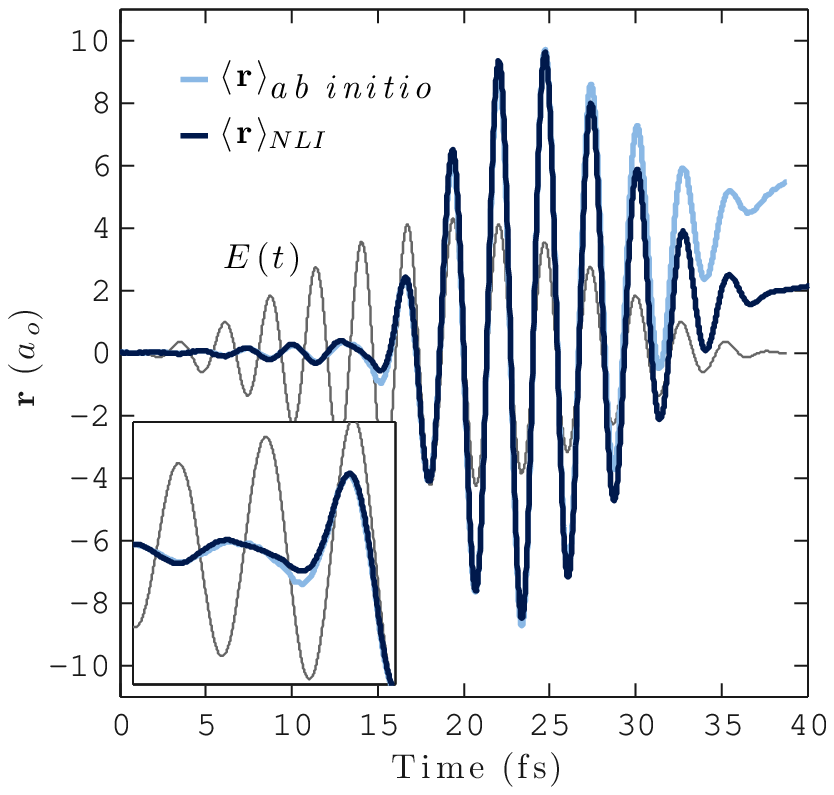}
\caption{The predicted \textit{ab initio} Hydrogen and NLI dipole moments are compared.}
\label{fig:dipole}
\end{figure}

The last quantity compared with the \textit{ab initio} simulation is the dipole moment, $\mathbf d(t)$ ($-\langle \mathbf r \rangle$), shown in Fig.\ref{fig:dipole}.  Again, reasonably good agreement is observed.  One feature of interest occurs at approximately 13 fs as shown in the inset, at which point $\mathbf r(t)$ changes relative phase with the applied field, $\mathbf E(t)$.  Prior to 13 fs, both plots of $\langle \mathbf r \rangle$ are seen to be out of phase with the electric field, and afterwards largely in phase.  This can be understood in the following way: the ``bound'' electron response is largely out of phase with the field and initially dominates.  As $\mathbf E(t)$ increases in strength, some of the electron wavefunction is excited to continuum states, leaving the vicinity of the ion and contributing as a ``free'' response to the dipole moment.  Because the spatial excursions of the free wavefunction are large compared to the wavefunction in the ground state, a comparatively small fraction of unbound wavefunction will dominate the atomic dipole, causing the net dipole moment to change sign with respect to the field. 

\subsection{PMPB ionization Theory Comparison}
In this section we compare the ionization rate predicted by the PMPB model with that of the NLI model as given by Eq.\eqref{wproxy}.  The PMPB ionization model predicts the probability of ionization of a Coulomb bound electron in the presence of a\ignore{ low frequency ($\omega_{laser} << |E_0|/\hbar$),} low field amplitude ($\mathrm{max}|\mathbf E(t)| < |E_0|/a_0$) sinusoidally varying electric field.  The rate shares the same exponential dependence on the electric field amplitude as the rate predicted by Keldysh, but includes an improved field-dependent Coulomb correction.  Direct comparison with the PMPB model is complicated by the fact that it only predicts a cycle averaged rate for monochromatic fields.  The time dependent simulations produced by the NLI model require that the field be zero on the semi-infinite range of $t < t_0$ (for arbitrary $t_0$), thus a true monochromatic field cannot be realized.  Thought was given to the most appropriate choice of the envelope for each of the parameter scans below, as will be discussed.

\subsubsection{FREQUENCY DEPENDENCE}
\begin{figure}
\centering
{\includegraphics[width=8cm]{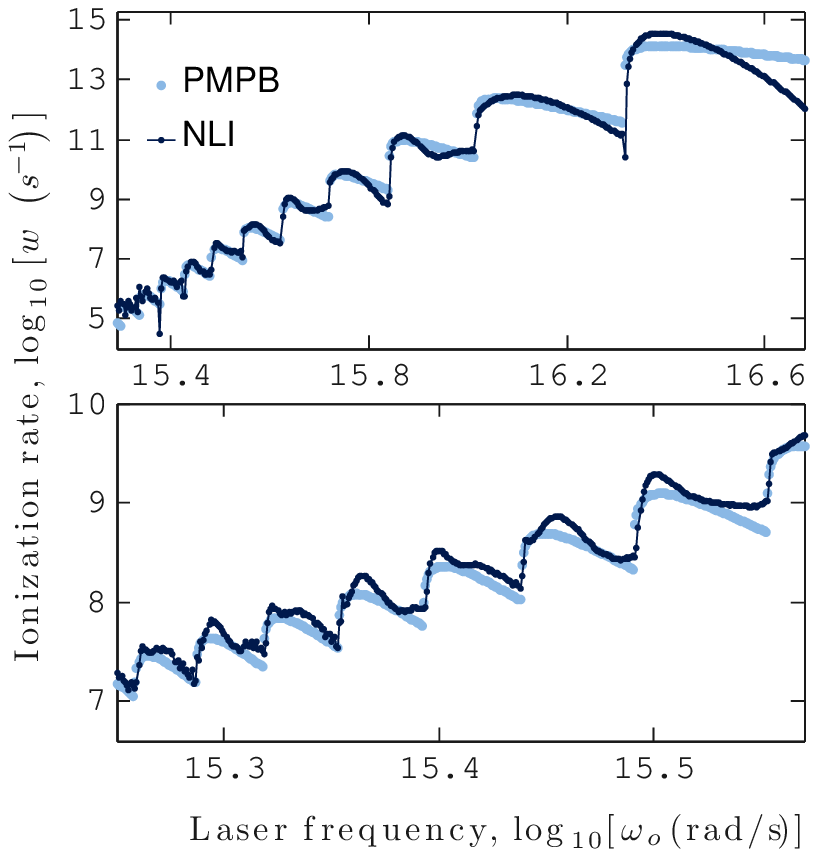}}
\caption{Top: the PMPB ionization rate for a monochromatic laser field is compared with the NLI rate.  The NLI rate is obtained by averaging Eq.\eqref{wproxy} over a the constant envelope region of laser pulse with I$_0 = 1\times 10^{13}$ W/cm$^2$. Bottom: the same result for optical frequencies, with the laser intensity increased to I$_0 = 1.9 \times 10^{13}$ W/cm$^2$.}
\label{fig:FREQ}
\end{figure}

Comparison of the frequency dependence is a crucial test for the NLI model.  The PMPB model predicts a strong dependence of ionization rate on the laser frequency, with local maxima in the rate $w$ occurring for each N-photon resonance, when $N \hbar \omega_{laser} \approx |E_0|$.  This expression is only approximate because the laser field distorts the effective binding potential energy (i.e., they are AC stark shifted 
\citep{Delone1999}).  To compare the frequency dependence of the PMPB and NLI ionization rates, a long constant amplitude pulse was used.  The envelope is piecewise defined to have a 15 fs $\sin^2(t)$ ramp to a constant amplitude of 1e13 W/cm$^2$ for 55 fs before symmetrically ramping back down to zero.  The profile is modulated by a carrier frequency, and subsequent ionization rates are calculated for electrons initialized in the ground state using the PMPB and NLI models.  
Fig.\ref{fig:FREQ} shows the NLI rate averaged over an integer number of laser cycles between 25 and 50 fs and plotted against the PMPB rate for the same field amplitude and frequency.  The NLI model is seen to reproduce each N-photon resonance predicted by the PMPB model; the highest peak occurs at the single photon ionization rate, where $\hbar \omega_{\text{laser}} \approx 1.2 |E_0|$.  Above this frequency, the ionization rate drops off as the electron cannot respond quickly enough to the laser field oscillations.  On increasing the intensity of the laser to 1.9$\times 10^{13}$ W/cm$^2$, the procedure was repeated for typical laboratory laser frequencies, shown in Fig.\ref{fig:FREQ}.
\begin{figure}
\centering
\end{figure}
\subsubsection{INTENSITY DEPENDENCE}
\begin{figure}
\includegraphics[width=8cm]{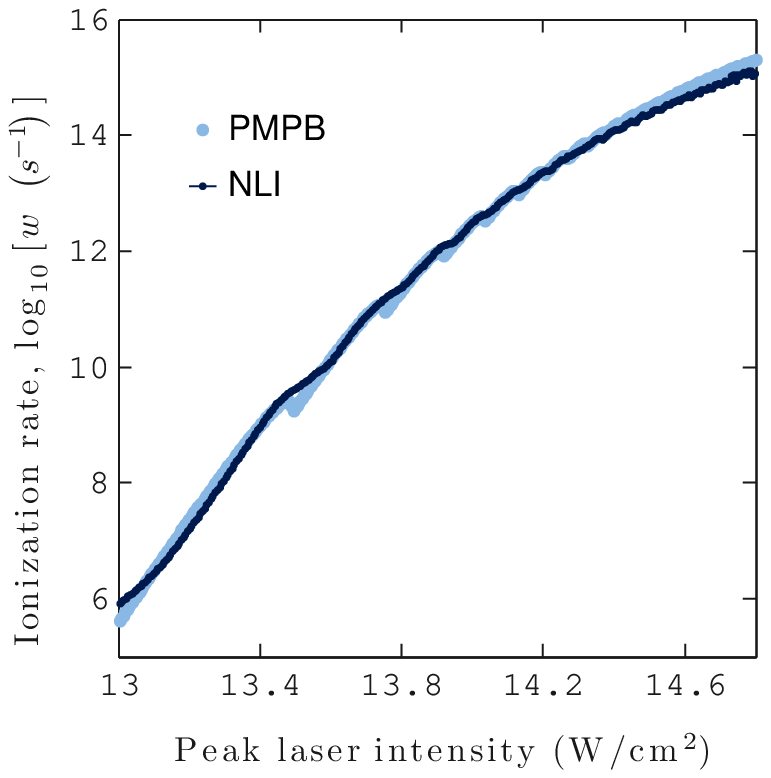}
\caption{Intensity dependence of the NLI (pulse averaged) ionization rate compared with the (pulse averaged) PMPB rate for a 14.1 fs, 800 nm pulse.}
\label{fig:RateVsIntensity}
\end{figure}

The intensity dependence of the ionization rate was also examined.  While it would be preferable to repeat the procedure used to determine the frequency dependence of the ionization rate, $w$ varies so strongly with intensity that this method must be modified.  Over a relatively modest range of intensity variation, the bound electron probability varies from an undetectable drop to (near) complete electron probability depletion.  Instead, a technique is used that measures the effective ionization rate for the entire laser pulse.  In this scenario, a 14.1 fs pulse of 800 nm light impinges on the Hydrogen atom.  The effective ionization rate is given in terms of the total drop in bound probability and the intensity full-width half-max duration of the pulse:
\begin{equation}
w_{\text{eff}} \equiv -\frac{1}{T_{\text{fwhm}}} \ln\left[\tilde \rho_b(t_f)\right].\label{weff}
\end{equation}
The same quantity can be determined for the PMPB model by first solving for $\rho_b(t)$ via Eq.\eqref{bprob} and then inserting into Eq.\eqref{weff}.  This procedure was repeated for various peak laser intensities, and the result plotted in Fig.\ref{fig:RateVsIntensity}.  This method demonstrates good agreement between the PMPB and NLI models over a wide range of intensities.  At the lowest laser intensities plotted, the ionization rate is artificially increased due to small numerical damping of the bound probability as measured by $\tilde \rho_b(t)$.   At the highest intensities, accuracy is limited by the effect of depletion: at 1x10$^{15}$ W/cm$^2$, the electron is completely ionized before the end of the pulse.  Since T$_{\text{fwhm}}$ is unchanged, the resulting effective rate is artificially suppressed.

\section{Extensions of the NLI model}\label{sec:extensions}
One of the noticeable omissions of the NLI model, as explored in this paper, is the existence of multiple bound states.  While the single state model is appropriate for many systems in the single active electron regime, there is utility in including multiple bound states; it has been suggested, for example, that the presence of additional bound states can have an effect on the non-linear polarizability of atoms \citep{Andreasen2013}.  The NLI model explored in this paper can be extended to include an arbitrary number of bound states using the following substitution:
\begin{subequations}\label{extension}
\begin{align}
Vu(\mathbf r)S(t) & \to \sum_i^N V_i u_i(\mathbf r)S_i(t),\\
S_i(t) &\to  \! \int \! \! \mathrm{d^3 \mathbf r}  \ u_i(\mathbf r)\psi(\mathbf r,t) \label{Ss},
\end{align}
\end{subequations}
where $u_i(\mathbf r)$ represents a set of orthogonal basis functions.  The NLI model then permits $N$ bound states with eigenvalues determined by the associated $V_i$.  A natural choice of basis functions is the three dimensional Gaussian-Hermite polynomials of which the $u(\mathbf r)$ used in this paper is the first,\ignore{
\begin{equation}
u_{i,j,k}(\mathbf r) = x^i y^j z^ke^{-x_i^2/2\sigma_i^2 - x_j^2/2\sigma_j^2 -x_k^2/2\sigma_k^2},
\end{equation}}
but other choices are possible.  Multiple bound states will be explored in future work.
\section{Conclusions}
This paper examines a phenomenological nonlocal potential model that provides an efficient method for calculating the atomic dipole in the presence of a laser electric field.  When compared to an \textit{ab initio} simulation of atomic Hydrogen, the NLI model gives surprisingly accurate results for the bound probability and atomic dipole moment.  The ionization rate for extended ranges of laser frequency and intensity were measured by comparing to the PMPB ionization rate, reproducing results again with surprising accuracy.  The extremely low computational overhead (when compared to full TDSE simulations) coupled with time dependent quantum dynamics of the atomic response make the NLI model a promising tool for calculating the nonlinear polarization in laser propagation simulations.

\section*{Acknowledgements}
The authors would like to acknowledge L. Johnson for fruitful discussions.  This work was supported by funding from ONR and DOE.

\section{Appendix}
Here we provide a sketch of the method used to treat the infinite integral in Eq.\eqref{Sgen}.  We begin by representing the integral equation for $S(t)$ schematically as:
\begin{equation}\label{Sk}
S(t)  = \int \limits_{-\infty}^t \mathrm{dt'}  K(t,t') S(t'),
\end{equation}
where $\mathbf E(t) = 0$ for $t<0$.  We would like to make use of the fact that the integral contribution over the infinite past can be expressed analytically in the absence of $E(t)$:
\begin{equation}\label{I1value}
\begin{split}
&S(t) = \int \limits_{-\infty}^0 \mathrm{dt'} K_0(t,t')S(t') = \\
&\frac{2^{5/2} V}{\sqrt{2+it}}\left[ 1 - \sqrt{(2+it)\pi|E_0|}\mathrm{erfc} \left(\sqrt{(2+it)|E_0|}\right)e^{(2+it)|E_0|} \right],
\end{split}
\end{equation}
where $K_0(t,t')$ is the field free kernel in Eq.\eqref{So}.  We rewrite the general form of $S(t)$ by splitting the integral as follows:
\begin{align}\label{splitkernel}
S(t)  &= \int \limits_{-\infty}^0 \mathrm{d}t'  K_1(t,t') S(t') + \int \limits_{0}^t \mathrm{d}t' K(t,t') S(t'),
\intertext{where}
\begin{split}
K_1(t,t') &= \frac{2^{3/2}iV}{{\left[ 2 + i (t-t') \right] }^\frac{3}{2}} \exp \left[ i\mathscr S(t;0) - \frac{1}{2}\mathbf r^2_0(t;0) \right] \times \\
&\exp\left[ \frac{1 + i(t-t')}{2+i(t-t')} \frac{|\mathbf r_0(t;0) - i \mathbf v_0(t;0)|^2}{2}  \right].
\end{split}
\end{align}
The trajectory variables $\mathbf v_0(t;0)$, $\mathbf r_0(t;0)$ in $K_1$ depend only on $t$, inviting the kernel to be written in terms of $K_0(t,t')$ as follows:

\begin{align}\label{ksplit}
\begin{split}
&K_1(t,t') =  f(t)K_0(t,t') \times \\
&1 + \left[\exp\left[ \left( \frac{1 + i(t-t')}{2+i(t-t')} -1 \right)\frac{|\mathbf r_0(t;0) - i \mathbf v_0(t;0)|}{2}  \right] -1 \right],
\end{split}
\intertext{where}
&f(t) \equiv \exp \left[ i\mathscr S(t;0) - \frac{1}{2}\mathbf r_0^2(t;0) + \frac{|\mathbf r_0(t;0) - i \mathbf v_0(t;0)|^2}{2} \right].
\end{align}
The advantage gained is that the first term in Eq.\eqref{ksplit} is solvable analytically via Eq.\eqref{I1value}, since $f(t)$ can be pulled out of the integral over $t'$.   Although the remaining terms (in square brackets) must be truncated and solved numerically, they vanish in the limit that $E(t) \to 0$ (i.e., $\mathbf v_0, \mathbf r_0 \to 0$) and as $(t-t') >> 1$.  Evaluating $S(t)$ this way permits a smooth transition from the analytic value of $S(t < 0)$ to the numerically calculated value for $S(t>0)$, whereas truncation of the integral in Eq.\eqref{Sk} creates a sharp discontinuity even for $E(t) = 0$.  
Finally, calculating the second term in Eq.\eqref{splitkernel} is straightforward, having obtained the history of $S(t)$ by the method outlined above. 
\ignore{
\subsection{Numerical Eigenvalue corrections}
While the relation between $E_0$ and $V$ given in Eq.\eqref{dispersion} is exact for the continuous system, the numerical system admits a slightly altered eigenvalue based on the step size.  The general equation for $S(t)$ is rewritten for discrete step size:
\begin{equation}\label{Snumerical}
S(n\Delta t') = \Delta t' \sum_{i=-\infty} ^{i=n} K(n\Delta t',i\Delta t')S(i\Delta t')
\end{equation}}

\bibliographystyle{apsrev.bst}
\bibliography{/Users/trensink/Dropbox/publication/Bibliographies/NLI_REFS}
\end{document}